\documentclass[aps,prd,twocolumn,superscriptaddress,nofootinbib,preprintnumbers]{revtex4-1}
\pdfoutput=1 
\usepackage{slashed}
\usepackage{color, verbatim}
\usepackage{latexsym}
\usepackage{amssymb}
\usepackage{amsmath}
\usepackage{graphicx}
\graphicspath{{figs/}}
\usepackage{arydshln}
\usepackage[dvipsnames]{xcolor}
\usepackage{adjustbox}
\usepackage{easybmat}
\usepackage{bbm}
\usepackage{bm}
\usepackage{adjustbox}
\usepackage{booktabs}
\usepackage{multirow}
\usepackage{rotating}
\usepackage{soul}
\usepackage{mathtools}
\DeclarePairedDelimiter{\bp}{\big(}{\big)}
\DeclareMathOperator{\Tr}{Tr}
\allowdisplaybreaks

\begin{document}

\preprint{FTUV-21-0416.5531}
\preprint{IFIC/21-09}

\title{Neutrino masses in the Standard Model effective field theory}

\author{Mikael Chala}
\email{mikael.chala@ugr.es}
\affiliation{Departamento de F\'isica Te\'orica y del Cosmos,
Universidad de Granada, E--18071 Granada, Spain}
\author{Arsenii Titov}
\email{arsenii.titov@ific.uv.es}
\affiliation{Departament de F\'isica Te\`orica, Universitat de Val\`encia 
and IFIC, Universitat de Val\`encia--CSIC,
Dr.~Moliner 50, E--46100 Burjassot, Spain}

\begin{abstract}
We compute the leading-logarithmic correction to the neutrino mass matrix in the Standard Model effective field theory (SMEFT) to dimension seven. In the limit of negligible lepton and down-type quark Yukawa couplings, it receives contributions from the Weinberg dimension-five operator as well as from 11 dimension-six and five dimension-seven independent interactions. Two of the main implications we derive from this result are the following. First, we find dimension-seven operators which, despite violating lepton number, do not renormalise neutrino masses at one loop. And second, we demonstrate that the presence of dimension-six operators around the TeV scale can modify the Standard Model prediction by up to $\mathcal{O}(50\,\%)$. Our result comprises also one step forward towards the renormalisation of the SMEFT to order $v^3/\Lambda^3$.
\end{abstract}

\maketitle

\section{Introduction}
\label{sec:intro}
%
The non-vanishing neutrino masses~\cite{Fukuda:1998mi,Toshito:2001dk,Giacomelli:2001td,Fukuda:2001nj,Fukuda:2001nk,Ahmad:2002jz,Ahmad:2002ka} remain still 
the only indisputable evidence of physics beyond the Standard Model (SM) 
of particle physics.
Their small value suggests that most likely the new physics threshold $\Lambda$ is well above the electroweak (EW) scale given by
$v\sim 246$~GeV~%
\footnote{Of course, neutrinos could be Dirac particles 
(as all other fermions in the SM), 
implying the existence of the right-handed fields $\nu_R$. 
Also, Majorana masses of the active left-handed neutrinos can be generated by relatively light new physics. Though these possibilities are definitely plausible, in the present article 
we assume that neutrino masses are generated by heavy new physics.}.
This in turn justifies using the SM effective field theory (SMEFT)~\cite{Brivio:2017vri} to describe particle physics at low energies.

If the baryon number is conserved~\cite{Zyla:2020zbs}, only SMEFT operators of an odd dimension contribute to the neutrino mass matrix $M_\nu$ at tree level~\cite{Kobach:2016ami}. Neglecting operators of dimension nine and higher, we have only:
\begin{equation}\label{eq:O5andO7}
 \begin{split}
 \mathcal{O}_{LH}^{(5)} &= \epsilon_{ij}\epsilon_{mn} \bp{L^i C L^m} H^j H^n\,, \\%
 \mathcal{O}_{LH}^{(7)} &= \epsilon_{ij}\epsilon_{mn} \bp{L^i C L^m} H^j H^n \bp{H^\dagger H}\,.
 \end{split}
\end{equation}
We will refer to both of them as Weinberg operators~\cite{Weinberg:1979sa}.
Here, $L$ and $H$ stand for the left-handed lepton and the Higgs doublets, respectively, $\epsilon$ is the totally antisymmetric tensor, and $C$ denotes the Dirac charge conjugation matrix~\footnote{Here and in what follows, 
$\psi C \chi$ stands for $\overline{\psi^c}\chi = \psi^T C \chi$, where 
$\psi^c \equiv C \overline{\psi}^T$. 
We omit the transposition on $\psi$ in the fermion bilinears to lighten the notation (cf. Ref.~\cite{Liao:2016hru}).}.

These operators give the simple tree-level relation:
\begin{equation}\label{eq:treemass}
 M_\nu^\text{tree} = -\frac{v^2}{\Lambda}\left(\alpha_{LH}^{(5)} + \alpha_{LH}^{(7)}\frac{v^2}{2\Lambda^2}\right)\,,
\end{equation}
where $\alpha_{LH}^{(5)}$ and $\alpha_{LH}^{(7)}$ stand for the corresponding Wilson coefficients.
However, things are significantly different at the quantum level. In this article, we will be interested in computing the one-loop leading-logarithmic correction to Eq.~\eqref{eq:treemass}.
There are various motivations to address this exercise:

1.~The Wilson coefficients
in Eq.~\eqref{eq:treemass} can, at least partially, cancel each other. Such cancellations are feasible, as $\mathcal{O}_{LH}^{(5)}$ arises at a higher-loop order than $\mathcal{O}_{LH}^{(7)}$ in some ultraviolet (UV) completions of the SM~\cite{Babu:2009aq,Anamiati:2018cuq}. This implies that the relative size of the one-loop correction to $M_\nu^\text{tree}$ can be arbitrarily large.

2.~Radiative corrections are triggered, in particular, by dimension-six terms, proportionally to $\alpha_{LH}^{(5)}$. Consequently, accurate measurements of $M_\nu$ performed at different energy scales~\cite{Fuks:2020zbm} could provide new tests of dimension-six  interactions.

3.~Most importantly, this computation paves the way for renormalising the SMEFT to order $v^3/\Lambda^3$. In particular, we will have to compute the divergences of a number of dimension-six operators, which despite being calculated in Refs.~\cite{Jenkins:2013zja,Jenkins:2013wua,Alonso:2013hga} are not provided there explicitly. We will emphasise that it is only on the basis of this information that one can build on previous results for renormalising the SMEFT to higher orders~\cite{Criado:2018sdb,Chala:2020wvs}.

In what follows, we will neglect the small down-type quark and lepton Yukawa couplings. 
This approximation simplifies the calculation enormously, in particular, because operators with different lepton chiralities do not renormalise each other.

\section{Technical details}
%
We use the following form of the SM dimension-four Lagrangian:
\begin{align}
 \mathcal{L}_{\text{ren}} &= -\frac{1}{4}G_{\mu\nu}^A G^{A \mu\nu} -\frac{1}{4} 
W_{\mu\nu}^I W^{I \mu\nu} -\frac{1}{4}B_{\mu\nu}B^{\mu\nu} \nonumber\\
 &\phantom{{}={}}+ \left(D_\mu H\right)^\dagger \left(D^\mu H\right) + \mu_H^2 
H^\dagger H -\frac{1}{2}\lambda_H \left(H^\dagger H\right)^2\nonumber\\
 &\phantom{{}={}}+i\left(\overline{Q}\slashed{D} Q + \overline{u}\slashed{D} u + \overline{d}\slashed{D} d + \overline{L}\slashed{D} L +\overline{e}\slashed{D} e \right)\,\nonumber\\
 &\phantom{{}={}}-\left(\overline{Q} Y_d H d + \overline{Q} Y_u \tilde{H}u + \overline{L} Y_e H e + \text{h.c.}\right).
\end{align}
Beyond the already defined $L$ and $H$, we have introduced $e$ for the right-handed leptons, and $Q$ and $u,d$ for the left- and the right-handed quarks, respectively. The $SU(2)_L\times U(1)_Y$ EW gauge bosons are denoted by $W$ and $B$, respectively; while $G$ represents the gluon. We have also defined $\tilde{H}=\epsilon H^*$. We work in the limit $Y_d, Y_e\to 0$ and assume $Y_u$ to be real and diagonal. Finally, we use the minus-sign convention for the covariant derivative: 
\begin{equation}
 D_\mu = \partial_\mu - i g_1 Y B_\mu - i g_2 \frac{\sigma^I}{2} W^I_\mu - i g_s\frac{\lambda^A}{2} G^A_\mu \,,
\end{equation}
with $g_1$, $g_2$ and $g_s$ representing the $U(1)_Y$, $SU(2)_L$ and $SU(3)_C$ gauge couplings, respectively. $Y$ stands for the hypercharge; $\sigma^I$ with $I=1,2,3$ denote 
the Pauli matrices; and $\lambda^A$ with $A=1,\dots,8$ are the Gell-Mann matrices. 

The effective Lagrangian 
involving higher-dimensional interactions reads
\begin{equation}
 \mathcal{L}_{\text{eff}} = \frac{1}{\Lambda} \alpha_{LH}^{(5)} \mathcal{O}_{LH}^{(5)} 
 + \frac{1}{\Lambda^2} \sum_i \alpha_{i} \mathcal{O}_i 
 + \frac{1}{\Lambda^3} \sum_j \alpha_{j} \mathcal{O}_j\,,
\end{equation}
where $i$~($j$) runs over all independent dimension-six (dimension-seven) interactions $\mathcal{O}_{i}$ ($\mathcal{O}_{j}$), 
and $\alpha_i$ ($\alpha_j$) denote the corresponding Wilson coefficients.

At dimension five, there is only the Weinberg operator~\cite{Weinberg:1979sa}.
As a minimal set of dimension-six operators we use the Warsaw basis~\cite{Grzadkowski:2010es}. For dimension-seven interactions we choose the basis of Ref.~\cite{Lehman:2014jma} 
(improved in Ref.~\cite{Liao:2016hru}), 
which in particular includes $\mathcal{O}_{LH}^{(7)}$.

At the technical level, large logarithmic corrections can be inferred from the running of the 
operators in Eq.~\eqref{eq:O5andO7}, which can be in turn read from the one-loop effective action. 
There are four different kinds of contributions to this:
\begin{figure}[t]
 \includegraphics[width=0.32\columnwidth]{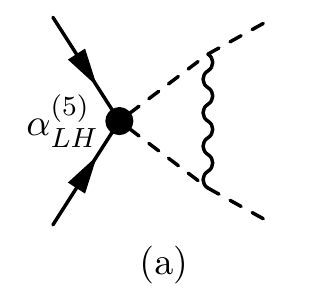}
 \includegraphics[width=0.32\columnwidth]{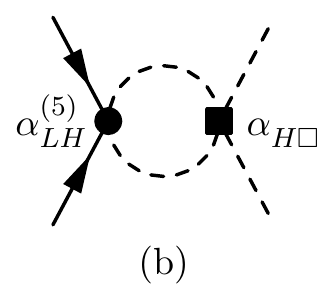}
 \includegraphics[width=0.32\columnwidth]{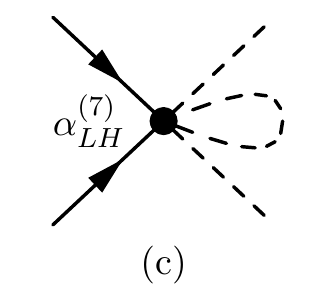}\\
 \includegraphics[width=0.32\columnwidth]{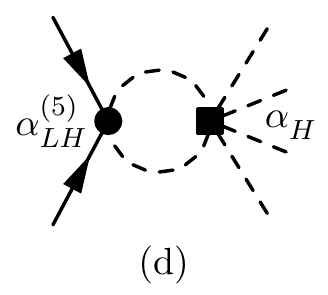}
 \includegraphics[width=0.32\columnwidth]{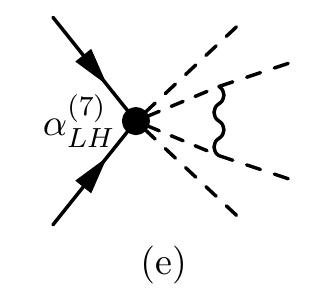}\\
 \includegraphics[width=0.32\columnwidth]{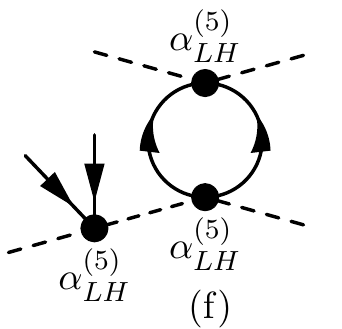}
 \includegraphics[width=0.48\columnwidth]{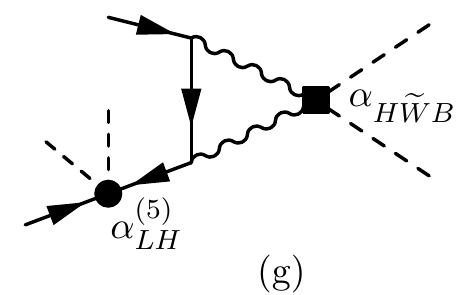}
 \caption{\it Examples of one-loop diagrams contributing directly to renormalisation of $\alpha_{LH}^{(5)}$ (upper row) and $\alpha_{LH}^{(7)}$ (middle row) 
 and indirectly via renormalisation of redundant dimension-six interactions (lower row). See the text for further details.}
 \label{fig:diagrams}
\end{figure}
\begin{table}[t]
 \setlength{\tabcolsep}{30pt}
 \begin{ruledtabular}
  \begin{tabular}{ll}
  $\mathcal{O}_{2B}$ & $-\frac{1}{2} \bp{\partial_\mu B^{\mu\nu}} \bp{\partial^\rho B_{\rho\nu}}$\\[0.1cm]
  $\mathcal{O}_{2W}$ & $-\frac{1}{2} \bp{D_\mu W^{I\mu\nu}} \bp{D^\rho W^{I}_{\rho\nu}}$\\[0.1cm]
  $\mathcal{O}_{BDH}$ & $\partial_\nu B^{\mu\nu} \bp{H^\dagger i \overleftrightarrow{D}_\mu H}$\\[0.1cm]
  $\mathcal{O}_{WDH}$& $D_\nu W^{I\mu\nu} \bp{H^\dagger i \overleftrightarrow{D}_\mu^I H}$\\[0.1cm]
  $\mathcal{O}_{DH}$ & $\bp{D_\mu D^\mu H}^\dagger \bp{D_\nu D^\nu H}$\\[0.1cm]
  $\mathcal{O}_{HD}^\prime$ & $\bp{H^\dagger H} \bp{D_\mu H}^\dagger \bp{D^\mu H}$\\[0.1cm]
  $\mathcal{O}_{HD}^{\prime\prime}$ & $\bp{H^\dagger H} D_\mu \bp{H^\dagger i \overleftrightarrow{D}^\mu H}$\\[0.1cm]
  $\mathcal{O}_{LD}$ & $\frac{i}{2} \overline{L} \left\{D_\mu D^\mu, \slashed{D}\right\}L$\\[0.1cm]
  $\mathcal{O}_{HL}^{\prime(1)}$ & $ \bp{H^\dagger H} \bp{\overline{L} i \overleftrightarrow{\slashed{D}} L}$\\[0.1cm]
  $\mathcal{O}_{HL}^{\prime\prime(1)}$ & $\partial_\mu \bp{H^\dagger H} \bp{\overline{L} \gamma^\mu L}$\\[0.1cm]
  $\mathcal{O}_{HL}^{\prime(3)}$ & $\bp{H^\dagger \sigma^I H} \bp{\overline{L} i \overleftrightarrow{\slashed{D}}^I L}$\\[0.1cm]
  $\mathcal{O}_{HL}^{\prime\prime(3)}$ & $D_\mu \bp{H^\dagger \sigma^I H} \bp{\overline{L} \gamma^\mu \sigma^I L}$\\[0.2cm]
  \hline\\[-0.3cm]
  $\mathcal{O}_{LHD}^{(R)}$ & $\epsilon_{ij}\epsilon_{mn} L^i C L^m H^j \square H^n$\\[0.1cm]
  \end{tabular}
 \end{ruledtabular}
 \caption{\it Relevant redundant operators of dimension six (top) and dimension seven (bottom).}
 \label{tab:redops6}
\end{table}

1.~We have direct renormalisation of the dimension-five Weinberg operator. There are diagrams with a single insertion of $\mathcal{O}_{LH}^{(5)}$ (see diagram~(a) in Fig.~\ref{fig:diagrams}), as well as diagrams involving both $\mathcal{O}_{LH}^{(5)}$ and a dimension-six interaction (see diagram~(b) in Fig.~\ref{fig:diagrams}) and diagrams with a single dimension-seven operator (see diagram~(c) in Fig.~\ref{fig:diagrams}). The latter two contribute proportionally to $\mu_H^2/\Lambda^2$. 
There are no diagrams with three insertions of $\mathcal{O}_{LH}^{(5)}$.

The mixing between operators of different dimensions, proportional to powers of $\mu_H/\Lambda$, makes this computation especially cumbersome. 
One could be tempted to neglect this for simplicity, 
as it is often done when using the dimension-six SMEFT at high energies, $E/v\gg 1$.
However, neutrino masses are proportional to $v\sim \mu_H$,
and hence dropping terms scaling as $\mu_H/\Lambda$ would be completely mistaken.

2.~We also have direct renormalisation of the dimension-seven Weinberg-like operator. There are diagrams involving $\mathcal{O}_{LH}^{(5)}$ and one dimension-six operator (see diagram~(d) in Fig.~\ref{fig:diagrams}), as well as diagrams with a single insertion of a dimension-seven term (see diagram~(e) in Fig.~\ref{fig:diagrams}).

3.~Then we also have renormalisation of (redundant) dimension-six interactions that, on shell (or in other words upon field redefinitions), contribute to the Weinberg operators (see diagrams (f) and (g) in Fig.~\ref{fig:diagrams}). A set of independent dimension-six redundant operators, namely an extension of the Warsaw basis to a Green basis, can be found in Ref.~\cite{Gherardi:2020det}. Neglecting lepton Yukawas, only operators involving neither $e$ nor quarks or the gluon can overlap on shell with the Weinberg operators, as the latter involve only left-handed leptons. It can be further checked that none of the operators in the class $\psi^2 X D$ is relevant. This leaves us with only the operators in the upper part of Tab.~\ref{tab:redops6}.
We have introduced 
$H^\dagger \overleftrightarrow{D}_\mu H \equiv H^\dagger D_\mu H - \left(D_\mu H\right)^\dagger H$ and 
$H^\dagger \overleftrightarrow{D}_\mu^I H \equiv H^\dagger \sigma^I D_\mu H - \left(D_\mu H\right)^\dagger \sigma^I H$. 
Flavour indices are suppressed to lighten the notation.

4.~Finally, we have renormalisation of redundant dimension-seven operators. Diagram (b) in Fig.~\ref{fig:diagrams} is representative of this case. Relative to our chosen basis of dimension-seven interactions, there is only one such independent redundant term, that we define in the lower part of Tab.~\ref{tab:redops6}.
Any other dimension-seven redundant interaction with non-zero overlapping with the Weinberg operators on shell can be reduced to $\mathcal{O}_{LHD}^{(R)}$ via algebraic identities or integration by parts. 

Within our bases of operators, divergences for the Weinberg interactions and for the aforementioned redundant operators can only be induced by the dimension-six and dimension-seven terms in Tabs.~\ref{tab:dim6} and \ref{tab:dim7}; and of course by the Weinberg operators themselves.
\begin{table}[t]
 \setlength{\tabcolsep}{30pt}
 \begin{ruledtabular}
  \begin{tabular}{ll}
   $\mathcal{O}_{H}$ & $\bp{H^\dagger H}^3$\\
   $\mathcal{O}_{H\square}$ & $\bp{H^\dagger H} \square \bp{H^\dagger H}$\\
   $\mathcal{O}_{HD}$ & $\bp{H^\dagger D^\mu H}^\ast \bp{H^\dagger D_\mu H}$\\
   $\color{gray}\mathcal{O}_{HB}$ &$\color{gray} \bp{H^\dagger H} B_{\mu\nu} B^{\mu\nu}$\\
   $\mathcal{O}_{HW}$ & $\bp{H^\dagger H} W_{\mu\nu}^I W^{I\mu\nu}$\,,\\
   $\color{gray}\mathcal{O}_{HWB}$ &$\color{gray}  \bp{H^\dagger \sigma^I H} W_{\mu\nu}^I B^{\mu\nu}$\\
  $\color{gray}\mathcal{O}_{3W}$ & $\color{gray} \epsilon^{IJK} W_{\mu}^{I\nu} W_{\nu}^{J\rho} W_{\rho}^{K\mu}$\\
  $\mathcal{O}_{H\widetilde{B}}$ & $\bp{H^\dagger H} \widetilde{B}_{\mu\nu} B^{\mu\nu}$\,,\\
  $\mathcal{O}_{H\widetilde{W}}$ & $\bp{H^\dagger H} \widetilde{W}_{\mu\nu}^I W^{I\mu\nu}$\\
  $\mathcal{O}_{H\widetilde{W}B}$ & $\bp{H^\dagger\sigma^I H} \widetilde{W}_{\mu\nu}^I B^{\mu\nu}$\\
  $\mathcal{O}_{uH}$ & $\bp{\overline{Q}\tilde{H} u} \bp{H^\dagger H}$\\
  $\mathcal{O}_{HL}^{(1)}$ & $\bp{H^\dagger i \overleftrightarrow{D}_\mu H} \bp{\overline{L}\gamma^\mu L}$\\
  $\mathcal{O}_{HL}^{(3)}$ & $\bp{H^\dagger i \overleftrightarrow{D}_\mu^I H} \bp{\overline{L} \gamma^\mu \sigma^I L}$\\
  $\color{gray}\mathcal{O}_{He}$ &$\color{gray} \bp{H^\dagger i\overleftrightarrow{D}_\mu H} \bp{\overline{e}\gamma^\mu e}$\\
  $\color{gray}\mathcal{O}_{HQ}^{(1)}$ &$\color{gray} \bp{H^\dagger i\overleftrightarrow{D}_\mu H} \bp{\overline{Q}\gamma^\mu Q}$\\
  $\mathcal{O}_{HQ}^{(3)}$ & $\bp{H^\dagger i\overleftrightarrow{D}_\mu^I H} \bp{\overline{Q} \gamma^\mu \sigma^I Q}$\\
  $\color{gray}\mathcal{O}_{Hu}$ &$\color{gray}\bp{H^\dagger i\overleftrightarrow{D}_\mu H} \bp{\overline{u}\gamma^\mu u}$\\
  $\color{gray}\mathcal{O}_{Hd}$ &$\color{gray}\bp{H^\dagger i\overleftrightarrow{D}_\mu H} \bp{\overline{d}\gamma^\mu d}$\\
  $\color{gray}\mathcal{O}_{LL}$ &$\color{gray} \bp{\overline{L}\gamma_\mu L} \bp{\overline{L}\gamma^\mu L}$
  \end{tabular}
 \end{ruledtabular}
 \caption{\it Dimension-six interactions that modify $M_\nu$ at one loop. Those in grey will be shown to give vanishing correction to Eq.~\eqref{eq:treemass} at the very end of the calculation.}
 \label{tab:dim6}
\end{table}

To fix the divergences of the relevant Wilson coefficients (that we name with a tilde hereafter), we compute all necessary one-particle-irreducible diagrams off shell. We proceed with the help of \texttt{FeynRules}~\cite{Alloul:2013bka}, \texttt{FeynArts}~\cite{Hahn:2000kx} and \texttt{FormCalc}~\cite{Hahn:1998yk}. We work in the background-field method in the Feynman gauge and use dimensional regularisation with space-time dimension $d=4-2\epsilon$. 

For the lepton number violating (LNV) operators we obtain:
\begin{widetext}
 \begin{align}
  \bp{\widetilde{\alpha}_{LH}^{(5)}}^{pq} = - \frac{1}{64\pi^2\epsilon} \bigg\{&\left(g_1^2 - 3g_2^2 + 4\lambda_H\right) \bp{\alpha_{LH}^{(5)}}^{pq} 
  -16 \bp{\alpha_{LH}^{(5)}}^{pq} \alpha_{H\square} \frac{\mu_H^2}{\Lambda^2} \nonumber \\
  &-8 \left[\bp{\alpha_{LH}^{(5)}}^{pr} \bp{\alpha_{HL}^{(1)}}^{rq} 
  -2 \bp{\alpha_{LH}^{(5)}}^{pr} \bp{\alpha_{HL}^{(3)}}^{rq} + p \leftrightarrow q\right] \frac{\mu_H^2}{\Lambda^2} 
  -16 \bp{\alpha_{LH}^{(7)}}^{pq} \frac{\mu_H^2}{\Lambda^2} \bigg\}\,, \\[0.2cm]
  \bp{\widetilde{\alpha}_{LH}^{(7)}}^{pq} = \frac{1}{256\pi^2\epsilon} \bigg\{&
  16\bp{\alpha_{LH}^{(5)}}^{pq} \left[ 
  6\alpha_H - 12\lambda_H\alpha_{H\square} + 2\lambda_H\alpha_{HD} 
  - 3g_1^2\alpha_{HB} - 3g_2^2\alpha_{HW} - 3g_1g_2\alpha_{HWB}\right]\nonumber \\
  &+24\left(g_1^2 + g_2^2\right) 
  \left[\bp{\alpha_{LH}^{(5)}}^{pr} \bp{\alpha_{HL}^{(1)}}^{rq} - \bp{\alpha_{LH}^{(5)}}^{pr} \bp{\alpha_{HL}^{(3)}}^{rq} + p \leftrightarrow q
	\right] \nonumber \\ 
  &-32 \lambda_H  
  \left[\bp{\alpha_{LH}^{(5)}}^{pr} \bp{\alpha_{HL}^{(1)}}^{rq} -2 \bp{\alpha_{LH}^{(5)}}^{pr} \bp{\alpha_{HL}^{(3)}}^{rq} + p \leftrightarrow q\right] \nonumber \\
  &+ 8\left(3g_2^2 - 20\lambda_H\right) \bp{\alpha_{LH}^{(7)}}^{pq}
  + 6g_2^3 \left[ 4\alpha_{LHW}^{pq} + g_2\alpha_{LHD1}^{pq} + p \leftrightarrow q\right] \nonumber \\
  &+3\left(g_1^4 + 3g_2^4 + 2g_1^2g_2^2\right) 
 	\left[\alpha_{LHD2}^{pq} + p \leftrightarrow q\right]
	+48 Y_u^{rt} Y_{u}^{us} Y_u^{ut} \left[\alpha_{QuLLH}^{rspq} + p\leftrightarrow q\right]
 	\bigg\}\,,\label{eq:divLH7}\\[0.2cm]
  \bp{\widetilde{\alpha}_{LHD}^{(R)}}^{pq}= \frac{1}{128\pi^2\epsilon} \bigg\{&16 \bp{\alpha_{LH}^{(5)}}^{pq} \alpha_{H\square} 
  -16 \left[\bp{\alpha_{LH}^{(5)}}^{pr} \bp{\alpha_{HL}^{(1)}}^{rq} 
  -2 \bp{\alpha_{LH}^{(5)}}^{pr} \bp{\alpha_{HL}^{(3)}}^{rq} + p \leftrightarrow q\right] \nonumber \\
  &+3\left[2 g_2^2 \alpha_{LHD1}^{pq} + g_2^2 \alpha_{LHD2}^{pq}
  +4 Y_u^{rs} \alpha_{QuLLH}^{rspq}  + p \leftrightarrow q\right]
  \bigg\}\,,\label{eq:divLHDR}
 \end{align}
\end{widetext}
where the sum over repeated flavour indices $r,s,t,u$ is implied. The pieces proportional to dimension-seven operators in Eqs.~\eqref{eq:divLH7} and \eqref{eq:divLHDR}  were partially computed for the first time in Ref.~\cite{Liao:2019tep} (see also Ref.~\cite{Cirigliano:2017djv}), which we have used for cross-check. 
It is worth noting that the renormalisation of the Weinberg operators by $\mathcal{O}_{LHW}$ 
has been previously studied in~\cite{Davidson:2005cs,Bell:2006wi}.
\begin{table}[t]
 \setlength{\tabcolsep}{20pt}
 \begin{ruledtabular}
  \begin{tabular}{ll}
  $\mathcal{O}_{LHD1}$ & $\epsilon_{ij} \epsilon_{mn} \bp{L^i C D^\mu L^j} H^m \bp{D_\mu H^n}$\\
  $\mathcal{O}_{LHD2}$ & $\epsilon_{im} \epsilon_{jn} \bp{L^i C D^\mu L^j} H^m \bp{D_\mu H^n}$\\
  $\mathcal{O}_{LHW}$ & $\epsilon_{ij} (\epsilon\sigma)_{mn} \bp{L^i C \sigma_{\mu\nu} L^m} H^j H^n W^{I\mu\nu}$\\
  $\mathcal{O}_{QuLLH}$ & $\epsilon_{ij} \bp{\overline{Q}u} \bp{LCL^i} H^j$
 \end{tabular}
 \end{ruledtabular}
 \caption{\it Dimension-seven interactions that modify $M_\nu$ at one loop.}
 \label{tab:dim7}
\end{table}

The divergences of the redundant dimension-six operators read:
\begin{align}\label{eq:red6div1}
 \widetilde{\alpha}_{2B} &= \widetilde{\alpha}_{DH} = \widetilde{\alpha}_{LD} = 0\,,\\
 \widetilde{\alpha}_{2W} &= -\frac{3g_2}{4\pi^2\epsilon}\alpha_{3W}\,,\\\nonumber
 \widetilde{\alpha}_{BDH} &= -\frac{g_1}{96\pi^2\epsilon} \bigg[\alpha_{H\square} + \alpha_{HD} -4\alpha_{He}^{pp} - 4\bp{\alpha_{HL}^{(1)}}^{pp} \\
 &+8\alpha_{Hu}^{pp} - 4\alpha_{Hd}^{pp} + 4\bp{\alpha_{HQ}^{(1)}}^{pp}\bigg]\,,\\\nonumber
 \widetilde{\alpha}_{WDH} &= -\frac{g_2}{96\pi^2\epsilon}\bigg[\alpha_{H\square} + 18g_2\alpha_{3W} \\
 &+4\bp{\alpha_{HL}^{(3)}}^{pp}+12\bp{\alpha_{HQ}^{(3)}}^{pp}\bigg]\,,\\\nonumber
 \widetilde{\alpha}_{HD}' &= -\frac{1}{32\pi^2\epsilon}\bigg[6\left( g_1^2\alpha_{HB}+g_1 g_2\alpha_{HWB} + 3g_2^2\alpha_{HW}\right)\\\nonumber
 &+\left(3g_1^2-3g_2^2+2\lambda_H\right)\alpha_{HD} + \left(6g_2^2-4\lambda_H\right)\alpha_{H\square}\\\nonumber
 &+6 Y_u^{pq} \left(\alpha_{uH}^{pq} + \alpha_{uH}^{pq\ast}\right) 
 -24 Y_u^{pr} Y_u^{qr} \bp{\alpha_{HQ}^{(3)}}^{pq} \nonumber\\
 &-8\bp{\alpha_{LH}^{(5)}}^{pq}\bp{\alpha_{LH}^{(5)}}_{pq}^*\bigg]\,,\\
 \widetilde{\alpha}_{HD}'' & =-\frac{3i}{32\pi^2\epsilon} Y_u^{pq} \left(\alpha_{uH}^{pq}-\alpha_{uH}^{pq*}\right)\,,\\
 \bp{\widetilde{\alpha}_{HL}^{\prime(1)}}^{pq} &= \frac{3}{16\pi^2\epsilon} 
\bp{\alpha_{LH}^{(5)}}_{pr}^\ast \bp{\alpha_{LH}^{(5)}}^{rq}\,,\\
 \bp{\widetilde{\alpha}_{HL}^{\prime(3)}}^{pq} &= -\frac{1}{8\pi^2\epsilon} 
\bp{\alpha_{LH}^{(5)}}_{pr}^\ast \bp{\alpha_{LH}^{(5)}}^{rq}\,, \\
\bp{\widetilde{\alpha}_{HL}^{\prime\prime(1)}}^{pq} &= \frac{3}{32\pi^2\epsilon} \left(g_1^2 \alpha_{H\widetilde{B}} + 3g_2^2 \alpha_{H\widetilde{W}}\right) \delta^{pq}\,, \\
\bp{\widetilde{\alpha}_{HL}^{\prime\prime(3)}}^{pq} &= -\frac{3}{32\pi^2\epsilon} g_1 g_2 \alpha_{H\widetilde{W}B} \delta^{pq} \,.\label{eq:red6div2}
\end{align}
These terms are needed for computing one-loop corrections to $M_\nu$ to order $v^3/\Lambda^3$. More generally, renormalising the SMEFT to certain dimension requires knowledge about the counterterms of redundant operators at all smaller dimensions.

It is therefore important that divergences of redundant operators are given explicitly rather than only used in intermediate steps of the calculation. In this vein, it would be desirable if Eqs.~\eqref{eq:red6div1}--\eqref{eq:red6div2} are extended elsewhere to include the divergences of all redundant dimension-six terms.

Upon using the SMEFT equations of motion to $\mathcal{O}(1/\Lambda)$~\cite{Barzinji:2018xvu} for $B$, $W$, $H$ and $L$,
\begin{align}
 \partial^\nu B_{\mu\nu} &= \frac{g_1}{2}H^\dagger i\overleftrightarrow{D}_\mu H + g_1 Y^f\overline{f}\gamma_\mu f\,,\\[0.2cm]
 D^\nu W_{\mu\nu}^I &= \frac{g_2}{2} \left(H^\dagger i \overleftrightarrow{D}_\mu^I H + \overline{L}\gamma_\mu \sigma^I L + \overline{Q}\gamma_\mu\sigma^I Q\right)\,,\\[0.2cm]\nonumber
 D^2 H^i &= \mu_H^2 H^i -\lambda_H \bp{H^\dagger H} H^i 
 +Y_u^{pq}\epsilon^{ij} \overline{Q_p^j} u_q\\
 &- \frac{\big(\alpha_{LH}^{(5)}\big)_{pq}^{*}}{\Lambda} \epsilon^{ij} 
\left[\overline{L^j_q} \bp{\widetilde{H}^T L^c_p} + \bp{\overline{L_q} \widetilde{H}} L^{c,j}_{p}\right],\\[0.2cm]
 i\slashed{D}L^i_q &= -2 \frac{\big(\alpha_{LH}^{(5)}\big)_{pq}^{*}}{\Lambda} \widetilde{H}^i \bp{\widetilde{H}^T L^c_p}\,,
\end{align}
(with $f$ running over all fermions), we obtain the following on-shell projections onto the Weinberg-like operators:
\begin{align}\label{eq:reduction1}
 \bp{\mathcal{O}_{LHD}^{(R)}}^{pq} &\supset \mu_H^2 \bp{\mathcal{O}_{LH}^{(5)}}^{pq} - \lambda_H \bp{\mathcal{O}_{LH}^{(7)}}^{pq}\,,\\[0.2cm]
 \mathcal{O}_{2W} &\supset\frac{g_2^2}{2\Lambda} \bp{\alpha_{LH}^{(5)}}^{pq} \bp{\mathcal{O}_{LH}^{(7)}}^{pq} + \text{h.c.}\,,\\[0.2cm]
 \mathcal{O}_{WDH} &\supset -\frac{2g_2}{\Lambda} \bp{\alpha_{LH}^{(5)}}^{pq} \bp{\mathcal{O}_{LH}^{(7)}}^{pq} + \text{h.c.}\,,\\[0.2cm]
 \mathcal{O}_{HD}^\prime &\supset - \frac{1}{\Lambda} \bp{\alpha_{LH}^{(5)}}^{pq} \bp{\mathcal{O}_{LH}^{(7)}}^{pq} + \text{h.c.}\,,\\[0.2cm]
 \mathcal{O}_{HD}^{\prime\prime} &\supset -\frac{2i}{\Lambda} (\alpha_{LH}^{(5)})^{pq}(\mathcal{O}_{LH}^{(7)})^{pq}+\text{h.c.}\,,\\[0.2cm]
 \delta^{pq} \bp{\mathcal{O}_{HL}^{\prime\prime(1)}}^{pq} &\supset \frac{2i}{\Lambda}
 \bp{\alpha_{LH}^{(5)}}^{rp} \bp{\mathcal{O}_{LH}^{(7)}}^{rp} + \text{h.c.}\,,\\[0.2cm]
\delta^{pq} \bp{\mathcal{O}_{HL}^{\prime\prime(3)}}^{pq} &\supset -\frac{2i}{\Lambda} 
\bp{\alpha_{LH}^{(5)}}^{rp} \bp{\mathcal{O}_{LH}^{(7)}}^{rp} + \text{h.c.}\label{eq:reduction2}
\end{align}

On top of these quantities, the divergences of the Higgs and left-handed lepton kinetic terms, to order $1/\Lambda^2$, are also needed. We get:
\begin{align}
 K_H = - \frac{1}{32\pi^2\epsilon} \bigg[&g_1^2 + 3g_2^2 - 6 \Tr\left(Y_u^2\right) \nonumber \\
 &+ 2\left(2\alpha_{H\square} - \alpha_{HD}\right)\frac{\mu_H^2}{\Lambda^2}\bigg]\,,
\end{align}
as well as
\begin{equation}
 K_L = \frac{1}{64\pi^2\epsilon} \left(g_1^2 + 3g_2^2\right)\,.
\end{equation}

In order to derive the renormalisation group equations for 
the Wilson coefficients of the Weinberg operators, first we need to remove 
the operators which are redundant on shell. To this aim, we
use Eqs.~\eqref{eq:reduction1}--\eqref{eq:reduction2} 
to obtain $\widetilde{\alpha}_{LH}^{(5)}$ and $\widetilde{\alpha}_{LH}^{(7)}$ in the physical basis, which with a little abuse of notation we will still name the same.
Subsequently, we canonically normalise the kinetic terms by performing the perturbative field redefinitions $H\to (1 -K_H/2)H$ and $L\to (1 - K_L/2)L$.
The bare $\bp{\alpha_{LH}^{(5)}}^{0}$ is related to the renormalised counterpart by 
$\bp{\alpha_{LH}^{(5)}}^0 = \mu^{2\epsilon}Z_{LH}^{(5)}\alpha_{LH}^{(5)}$, with $Z_{LH}^{(5)} = 1-\widetilde{\alpha}_{LH}^{(5)}/\alpha_{LH}^{(5)}$. The running of $\alpha_{LH}^{(5)}$ is thus derived from the requirement that the bare term is independent of the renormalisation scale $\mu$, namely, $\mu\, \mathrm{d}\bp{\alpha_{LH}^{(5)}}^0/\mathrm{d}\mu = 0$. This gives:
\begin{equation}
 \mu\, \frac{\mathrm{d} \alpha_{LH}^{(5)}}{\mathrm{d}\mu} = \epsilon\,\alpha_{LH}^{(5)} \frac{\partial Z_{LH}^{(5)}}{\partial g_k} n_k g_k\,,
\end{equation}
with $g_k$ running over all Lagrangian couplings, renormalisable or not, and with $n_k$ representing the tree-level anomalous dimensions of $g_k$, namely, the values required to keep the
action dimensionless in $d=4-2\epsilon$ dimensions.
A completely analogous derivation holds of course for the dimension-seven Weinberg-like operator.

Solving the corresponding renormalisation group equations to the leading-logarithmic approximation, and upon EW symmetry breaking, we obtain:
\begin{widetext}
 \begin{align}
  \label{eq:loopmass}
   \delta{M_\nu^{pq}} &=-\frac{v^2}{16 \pi^2\Lambda}\log{\frac{\Lambda}{v}}
  \bigg\{
  \left[3g_2^2 - 2\lambda_H- 6 \Tr\left(Y_u^2\right)\right]    \big(\alpha_{LH}^{(5)}\big)^{pq} \nonumber \\
  &\phantom{{}={}}+\frac{v^2}{\Lambda^2} \big(\alpha_{LH}^{(5)}\big)^{pq} 
  \bigg[ 
  -4 \bp{\alpha_{LH}^{(5)}}^{st} \bp{\alpha_{LH}^{(5)}}_{st}^\ast 
  +6 \alpha_H
  +\frac{2}{3} \left(5g_2^2 - 12\lambda_H\right) \alpha_{H\square}
  +\frac{1}{2} \left(3g_1^2 - 3g_2^2 + 4\lambda_H\right) \alpha_{HD} 
  +6g_2^2 \alpha_{HW} \nonumber \\
  &\phantom{{}={}}
  +3 i g_1^2\alpha_{H\widetilde{B}} 
  +9 i g_2^2\alpha_{H\widetilde{W}} 
  +3 i g_1g_2\alpha_{H\widetilde{W}B} 
  +6 Y_u^{st} \alpha_{uH}^{st\ast}
  +\frac{4}{3} g_2^2 \bp{\alpha_{HL}^{(3)}}^{rr}
  +4 g_2^2 \bp{\alpha_{HQ}^{(3)}}^{rr} 
  -12 Y_u^{su} Y_u^{tu} \bp{\alpha_{HQ}^{(3)}}^{st}
  \bigg] \nonumber \\
  &\phantom{{}={}}+\frac{v^2}{\Lambda^2} 
  \frac{3}{2}\left(g_1^2+g_2^2\right)
  \bigg[\bp{\alpha_{LH}^{(5)}}^{pr} \bp{\alpha_{HL}^{(1)}}^{rq} - \bp{\alpha_{LH}^{(5)}}^{pr} \bp{\alpha_{HL}^{(3)}}^{rq} + p \leftrightarrow q\bigg] 
  +\frac{v^2}{\Lambda^2} \frac{3}{4}
  \bigg[g_1^2 + 5g_2^2 - 8\lambda_H - 8\Tr\left(Y_u^2\right)\bigg] \bp{\alpha_{LH}^{(7)}}^{pq} \nonumber \\
  &\phantom{{}={}}+\frac{v^2}{\Lambda^2}
  \bigg[\frac{3}{8} g_2^4 \alpha_{LHD1}^{pq} 
  +\frac{3}{16}\left(g_1^4 + 2 g_1^2 g_2^2 + 3 g_2^4\right) \alpha_{LHD2}^{pq} 
  +\frac{3}{2}g_2^3 \alpha_{LHW}^{pq} 
  +3 Y_u^{rt} Y_u^{us} Y_u^{ut} \alpha_{QuLLH}^{rspq}
  +p \leftrightarrow q\bigg]
 \bigg\} \,.
 \end{align}
\end{widetext}

At energies $\mu<v$, the neutrino mass matrix does not renormalise in the limit of negligible lepton masses~\cite{Jenkins:2017dyc}. Thus, Eq.~\eqref{eq:loopmass} comprises the leading correction to Eq.~\eqref{eq:treemass}, constituting the main result of this article.

Let us note that the renormalisation of the Weinberg dimension-five operator 
due to the SM interactions has been extensively studied a long time 
ago~\cite{Chankowski:1993tx,Babu:1993qv,Antusch:2001ck}. 
The first line in Eq.~\eqref{eq:loopmass} agrees with the previous result, 
which provides an additional check of our computation.

A more precise determination of $\delta M_\nu$ could be obtained upon accounting also for the running of the renormalisable couplings as well as the
dimension-six operators  themselves, which can be found in Refs.~\cite{Elias-Miro:2013gya,Elias-Miro:2013mua,Jenkins:2013zja,Jenkins:2013wua,Alonso:2013hga}.

\section{Discussion}
%
Equation~\eqref{eq:loopmass} becomes more important as smaller the tree-level neutrino mass is. The latter can be  approximately vanishing if, for example, both $\alpha_{LH}^{(5)}$ and $\alpha_{LH}^{(7)}$ are zero. In such a case, neutrino masses within the SMEFT are necessarily induced by dimension-seven operators 
(other than $\mathcal{O}_{LH}^{(7)}$), at a scale of order $\mathcal{O}(10^3)$~TeV for unit Wilson coefficients.

Not all combinations of the dimension-seven operators within our basis generate log-enhanced neutrino masses, though. For example, it can be directly read from Eq.~\eqref{eq:loopmass} that 
$\mathcal{O}=g_2\mathcal{O}_{LHW} - 4\mathcal{O}_{LHD1}$ 
gives $\delta M_\nu = 0$. (It is worth noting that we could only come to this result upon having included $\mu_H^2/\Lambda^2$ corrections in the renormalisation of $\alpha_{LH}^{(5)}$. Otherwise $\mathcal{O}$ would seem to induce a spurious non-vanishing $\delta M_\nu$ proportional to $\lambda_H$.)

This means that, within an UV completion of the SM leading to $\mathcal{O}$ (note that both $\mathcal{O}_{LHD1}$ and $\mathcal{O}_{LHW}$ can arise at tree level~\cite{Elgaard-Clausen:2017xkq}), neutrino masses would be finite at one-loop order up to leptonic Yukawa corrections. It would be interesting to investigate whether such finite piece vanishes or not.

Another case where Eq.~\eqref{eq:loopmass} becomes important is when both Weinberg operators cancel partially each other at tree level. As we anticipated in the Introduction, such cancellations are feasible, as $\mathcal{O}_{LH}^{(5)}$ can arise at higher-loop order than $\mathcal{O}_{LH}^{(7)}$ in UV completions of the SM~\cite{Babu:2009aq,Anamiati:2018cuq}.

\begin{figure}[t]
 \includegraphics[width=\columnwidth]{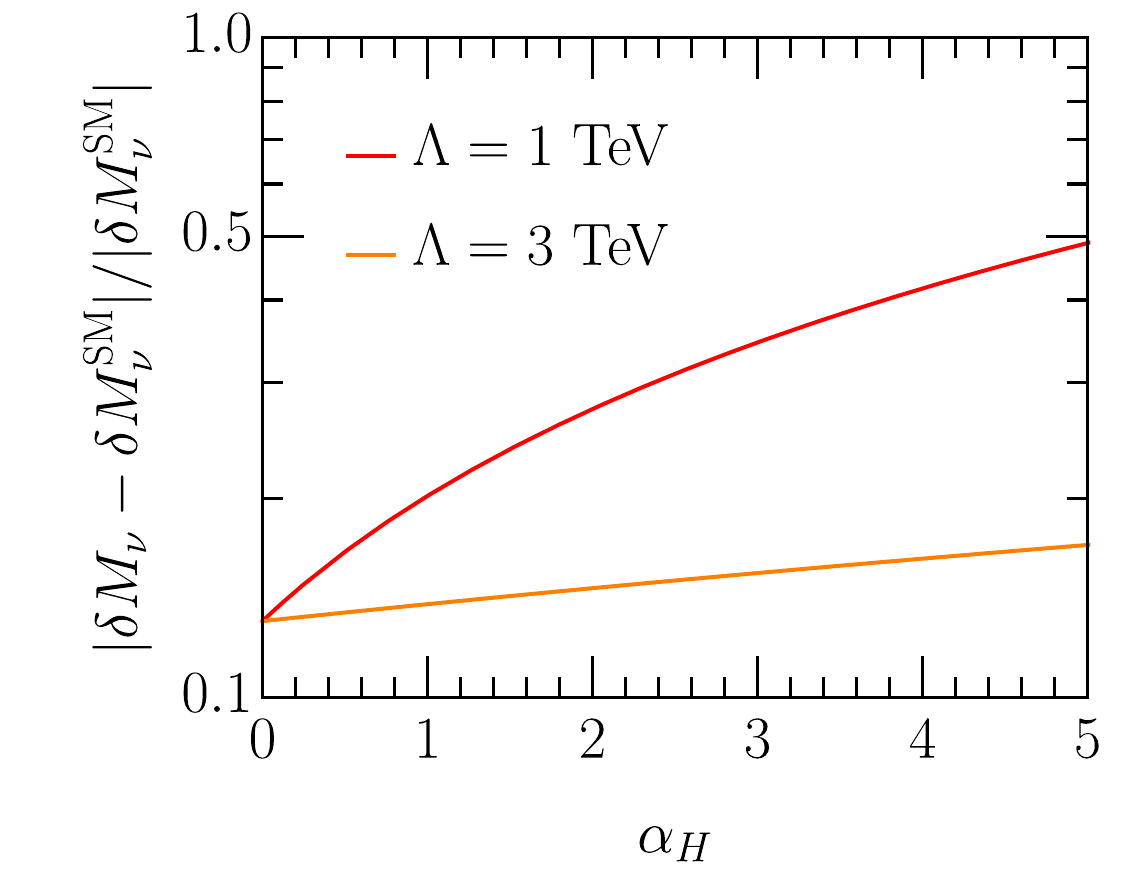}
 \caption{\it Impact of dimension-six interactions on the size of the leading-logarithmic %
correction to $M_\nu$ as a function of $\alpha_H$ for different values of the new physics scale $\Lambda$. 
The Wilson coefficients of other dimension-six operators 
have been set to their best-fit values from~\cite{Ellis:2020unq}. 
Dimension-seven operators have been assumed to vanish.
See the text for further details.}
 \label{fig:comparison}
\end{figure}

\begin{figure}[t]
 \includegraphics[width=\columnwidth]{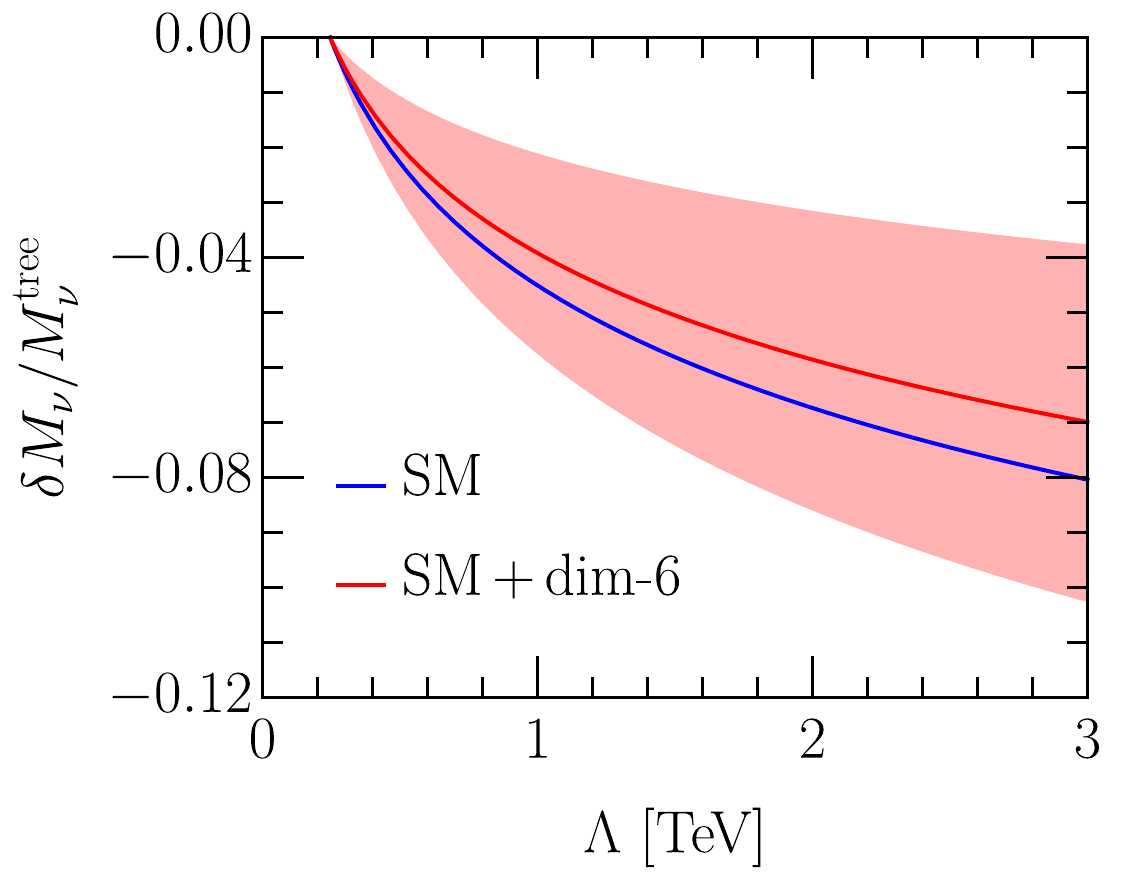} 
 \caption{\it Relative size of the leading-logarithmic correction to $M_\nu$ as a function of $\Lambda$, 
 with the Wilson coefficients of dimension-six operators 
 set to zero (blue line) and to their best-fit values from~\cite{Ellis:2020unq} (red line).
 The band represents the variation 
 of the Wilson coefficients of dimension-six operators
 across the 95\,\% confidence level ranges derived in~\cite{Ellis:2020unq}. 
Dimension-seven operators have been assumed to vanish.
 See the text for further details.}
 \label{fig:comparison2}
\end{figure}

Irrespective of potential cancellations, we find interesting analysing the importance of dimension-six operators in $\delta M_\nu$ relative to the SM contribution. To this aim, let us just work in the one-family case.
We compute $\delta M_\nu$ for different values of $\alpha_H$~\footnote{This Wilson coefficient is very weakly constrained by current data~\cite{DiVita:2017vrr,Chala:2018ari,Henning:2018kys}. Moreover, there are theoretical reasons to expect $\alpha_H/\Lambda^2\gtrsim 1$ TeV$^{-2}$~\cite{Chala:2018ari}, based on the possibility that the matter-antimatter asymmetry is induced by EW baryogenesis~\cite{Kuzmin:1985mm} due to a modified Higgs potential~\cite{Grojean:2004xa}.} assuming that all other dimension-six operators are fixed to their best (marginalised) fit values from Ref.~\cite{Ellis:2020unq} (other Wilson coefficients not included in Tab.~6 of that article, \textit{i.e.} $\alpha_{H\widetilde{B}}$, $\alpha_{H\widetilde{W}}$ and $\alpha_{H\widetilde{W}B}$ are set to zero because they are very constrained, 
see \textit{e.g.}~\cite{Cirigliano:2019vfc,Bernlochner:2018opw}; 
and dimension-seven operators are assumed to be zero for simplicity), and compare to the case where all dimension-six interactions vanish, $\delta M_\nu^\text{SM}$. 
We obtain that
$\lvert\delta M_\nu-\delta M_\nu^\text{SM}\rvert/\lvert\delta M_\nu^\text{SM}\rvert$ 
reaches $50\,\%$ for $\Lambda = 1$ TeV and $\alpha_H = 5$;
see Fig.~\ref{fig:comparison}.
On a different front, we compare $\delta M_\nu$ to $M_\nu^\text{tree}$ as a function of $\Lambda$ in Fig.~\ref{fig:comparison2}, this time allowing 
the Wilson coefficients of dimension-six operators 
to vary across the 95\,\% confidence level ranges~\cite{Ellis:2020unq}.
The plot does not go beyond $\Lambda > 3$ TeV, as some 
of these coefficients approach the non-perturbative limit in that regime.

It would be worth exploring further the impact of dimension-six corrections 
on the neutrino mass and mixing parameters. In particular, 
we believe that our results open the possibility of including neutrino data in the fits of the SMEFT.
For example, if $M_\nu$ is accurately measured at both low energies and near the TeV scale, Eq.~\eqref{eq:loopmass} can be used to constrain dimension-six interactions. Whether  $M_\nu$ can be precisely determined at colliders is still an open question, though. It depends significantly on how large the entries in $M_\nu$ are~\cite{Jenkins:2008ex,Fuks:2020zbm}.

Also, it would be particularly interesting to study to which extent the neutrino mass ordering is affected by the dimension-six  flavoured operators $O_{uH}$, $\mathcal{O}_{HL}^{(1)}$, $\mathcal{O}_{HL}^{(3)}$ and $\mathcal{O}_{HQ}^{(3)}$. For this matter, fits of the SMEFT to the data with relaxed (or none) flavour assumptions, as that in Ref.~\cite{Falkowski:2019hvp}, are needed.
In particular, it may happen that $M_\nu^{ee}$ (or better to say $\big(\alpha_{LH}^{(5)}\big)^{ee}$) 
vanishing at some high scale $\Lambda$ will be generated at the EW scale in the result of running 
induced by $\mathcal{O}_{HL}^{(1)}$ and $\mathcal{O}_{HL}^{(3)}$, cf.~Eq.~\eqref{eq:loopmass}. 
Consequently, neutrinoless double beta decay would have a non-zero rate. 
In the absence of the dimension-six interactions, this would not have been the case, 
since $\big(\alpha_{LH}^{(5)}\big)^{pq}$ runs proportionally to itself (in the considered approximation of negligible charged lepton Yukawa couplings).

Related to this discussion, it is also worth highlighting that several combinations of dimension-six operators do not modify $\delta M_\nu$. The most notable of these are: $\mathcal{O}_{3W}$, $\mathcal{O}_{HWB}$ and $g_2^2\mathcal{O}_H-\mathcal{O}_{HW}$; as well as $2g_2^2\mathcal{O}_H-3\mathcal{O}_{HQ}^{(3)}$ and $2\mathcal{O}_{HW}-3\mathcal{O}_{HQ}^{(3)}$ (for $Y_u\to 0$). It would be interesting to find deeper reasons, if any, behind these non-renormalisation results.

Finally, we would like to stress that, if the LNV and the dimension-six operators are suppressed by different scales, $\Lambda_\text{LNV}\gg\Lambda_\text{LNC}$, then the $v^3/(\Lambda_\text{LNV}\Lambda_\text{LNC}^2)$ piece of Eq.~\eqref{eq:loopmass} comprises the leading correction to the SMEFT renormalisation obtained in Refs.~\cite{Jenkins:2013zja,Jenkins:2013wua,Alonso:2013hga}; 
being, in particular, more important than the $v^2/\Lambda_\text{LNV}^2$ piece obtained in Ref.~\cite{Davidson:2018zuo}.
We plan to address the full renormalisation of the SMEFT to order $v^3/\Lambda^3$ in a future work.

\begin{acknowledgments}
We would like to thank Xiao-Dong Ma for help in understanding some of the computations in Ref.~\cite{Liao:2019tep}. We would like to thank Guilherme Guedes, Maria Ramos, Jose Santiago 
and Arcadi Santamaria for useful discussions. MC is supported by the Spanish MINECO under the Ram\'on y Cajal programme and partially by the
Ministry of Science and Innovation under grant number FPA2016-78220-C3-1/3-P (FEDER), SRA under grant PID2019-106087GB-C21/C22 (10.13039/501100011033) as well as by the Junta de Andaluc\'ia grants FQM 101, and A-FQM-211-UGR18 and P18-FR-4314 (FEDER). 
AT is supported by the “Generalitat Valenciana” under grant PROMETEO/2019/087 and by the FEDER/MCIyU-AEI grant FPA2017-84543-P.
\end{acknowledgments}

\bibliography{NuMassSMEFT}

\end{document}